\documentstyle[aps,pra]{revtex}
\begin{document}
\draft

\newcommand{\uu}[1]{\underline{#1}}
\newcommand{\pp}[1]{\phantom{#1}}
\newcommand{\be}{\begin{eqnarray}}
\newcommand{\ee}{\end{eqnarray}}
\newcommand{\ve}{\varepsilon}
\newcommand{\vs}{\varsigma}
\newcommand{\Tr}{{\,\rm Tr\,}}
\newcommand{\pol}{\frac{1}{2}}

\title{States of light via reducible quantization}

\author{Marek~Czachor}
\address{
Katedra Fizyki Teoretycznej i Metod Matematycznych\\ 
Politechnika Gda\'{n}ska, ul. Narutowicza 11/12, 80-952 Gda\'{n}sk,
Poland\\ 
email: mczachor@pg.gda.pl
}
\maketitle

\begin{abstract}
Multi-photon and coherent states of light are formulated in terms of a
reducible representation of canonical commutation relations. Standard properties of such states are
recovered as certain limiting cases. The new formalism leads to field
operators and not operator-valued distributions. The example of
radiation fields produced by a classical current shows an automatic
regularization of the infrared divergence.
\end{abstract}


\section{Introduction}

The present paper is a continuation of \cite{I} where a new approach to nonrelativistic quantum optics was proposed. The main idea of the new program of field quantization outlined in \cite{I} is to treat the frequencies of electromagnetic oscillators as observables (or their eigenvalues) and not as parameters. The most important implication of this single modification is a new representation of canonical commutation relations (CCR) which naturally occurs for field operators. The field operators are in this representation indeed operators and not operator-valued distributions and the formalism is much less singular than the traditional one.

The results discussed in \cite{I} were encouraging. However, before one will be able to say anything really conclusive about the role of the new quantization paradigm for the issue of infinities, one has to solve many intermediate problems. The list of open questions involves quantization of fermions, 
Poincar\'e covariant formalism, properties of multi-particle and coherent states, a full perturbative treatment of quantum electrodynamical 
$S$-matrix, and gauge invariance. 

The goal of the present paper is to give an explicit Poincar\'e covariant formulation of free Maxwell fields in the new representation, investigate the structure of multi-photon and coherent states of light, and properties of radiation fields in the simplest exactly solvable case. Some of the results are only outlined, for details the readers are refered to the preprints 
\cite{II,III,IV}. Ref. \cite{III} is devoted to free Dirac electrons and 
\cite{IV} gives a preliminary analysis of interactions in perturbation theory.

\section{Reducible quantization}

The mode quantization of free electromagnetic fields reduces to an
appropriate choice of the map
$
f(\bbox {k},s)\mapsto a(\bbox {k},s)
$
which replaces wave functions by operators. The idea of reducible
quantization is to take $a(\bbox {k},s)$ as an operator analogous to
the operator one finds for a harmonic oscillator whose frequency is
indefinite. In the nonrelativistic case there is only one parameter
$\omega$ and one finds \cite{I} the reducible representation of
CCR
$
a(\omega)=|\omega\rangle\langle\omega|\otimes a
$
where $a$ comes from the irreducible representation $[a,a^{\dag}]=1$ 
\cite{uwaga1}.
The electromagnetic field involves the 3-momentum $\bbox k$ and two
polarizations. Let $a_s$ be the annihilation operators corresponding
to an irreducible representation of the CCR algebra
$[a_s,a^{\dag}_{s'}]=\delta_{ss'}1$.  We accordingly define the 
{\it one-oscillator\/} indefinite-frequency annihilation
operators \cite{uwaga}
\be
a(\bbox k,s)
&=&
|\bbox k\rangle\langle \bbox k| \otimes a_s
\label{a(f)^+}
\ee
satisfying the commutation relations characteristic of a 
reducible representation of the CCR algebra
\be
[a(\bbox k,s),a(\bbox k',s')^{\dag}] &=&
\delta_{ss'}\delta_\Gamma(\bbox k, \bbox k')
\underbrace{|\bbox k\rangle\langle \bbox k| \otimes 1}_{I_{\bbox k}}
\ee 
Multi-oscillator fields are defined in terms of the reducible
representation which can be constructed from the single-oscillator
one as follows. Let $A$ be any operator acting in the
single-oscillator Hilbert space. If we denote by
$
A^{(n)}=\underbrace{I\otimes\dots\otimes I}_{n-1}
\otimes A\otimes \underbrace{I\otimes \dots\otimes I}_{N-n}
$
the extension of $A$ to the $N$-oscillator Hilbert space, then
\be
\uu a(\bbox k,s)
=
\frac{1}{\sqrt{N}}\sum_{n=1}^N
a(\bbox k,s)^{(n)}.\label{58}
\ee
Although the definition (\ref{58}) implies that the ``oscillators" which form the electromagnetic field are themselves of a bosonic type, one should not confuse them with photons. Photons, in this formalism, are quasi-particles corresponding to excitations of the oscillators. The operator 
$\uu a(\bbox k,s)$ removes one excitation from the ensemble, i.e. annihilates one photon. 

The factor 
$1/\sqrt{N}$ plays an important role for the thermodynamic limit $N\to\infty$. Its appearance can be given a physical interpretation if one considers a two-level system coupled to reducibly quantized vector potential. Since the coupling is linear, the interaction term becomes analogous to the one expressing a two-level system interacting with $N$ indefinite-frequency harmonic oscillators. As is widely known, a dual situation, representing an oscillator coupled to 
$N$ two-level systems (i.e. the Dicke model \cite{Allen}), also involves the factor  $1/\sqrt{N}$ in exactly the same place. Formally, the factor comes from the assumption that the density $N/V$ of two-level systems is constant, and is essential for the thermodynamic limit (cf. \cite{Lieb}). 

The reducible representation (\ref{58}) satisfies
\be
{[\uu a(\bbox k,s),\uu a(\bbox k',s')^{\dag}]} =
\delta_{ss'}\delta_\Gamma(\bbox k, \bbox k')
\uu I_{\bbox k}.\label{red-ccr}
\ee
The operator $\uu I_{\bbox k}$ at the right-hand-side of (\ref{red-ccr}) is given
by
$
\uu I_{\bbox k}
=
\frac{1}{N}\sum_{n=1}^N
I_{\bbox k}^{(n)}.
$
Its presence will influence orthogonality properties of multi-photon
states, as we shall see later. 
The electromagnetic field tensor and four-potential operators read
\be
{\uu F}_{ab}(x) &=&
\int d\Gamma(\bbox k)
\Bigg(
e_{ab}(\bbox k)
\Big(\uu a(\bbox k,-)e^{-ik\cdot x}
+\uu a(\bbox k,+)^{\dag}e^{ik\cdot x}\Big)+
\bar e_{ab}(\bbox k)
\Big(\uu a(\bbox k,-)^{\dag}e^{ik\cdot x}
+\uu a(\bbox k,+)e^{-ik\cdot x}\Big)\Bigg),\\
{\uu A}_a(x) &=& i\int d\Gamma(\bbox k)
\Bigg(
m_{a}(\bbox k)
\Big(\uu a(\bbox k,+)e^{-ik\cdot x}
-\uu a(\bbox k,-)^{\dag}e^{ik\cdot x}\Big) +
\bar m_{a}(\bbox k)
\Big(\uu a(\bbox k,-)e^{-ik\cdot x}
-\uu a(\bbox k,+)^{\dag}e^{ik\cdot x}\Big)
\Bigg).\label{A_a}
\ee
In the Penrose-Rindler spinor notation \cite{PR} one finds 
$k_a=\pi_A\bar\pi_{A'}$, $e_{ab}=\varepsilon_{A'B'}\pi_A\pi_B$,
$m_a=\omega_A\bar\pi_{A'}$, $\bar m_a=\pi_{A}\bar\omega_{A'}$, 
where $\omega_A\pi^A=1$. 
It is well known that field ``operators" of the standard theory are in fact operator-valued distributions. The reducibly quantized fields are more regular, as we shall see later.

\section{Action of the Poincar\'e group on field operators}

Denote, respectively, by $\Lambda$ and $y$ the $SL(2,C)$ and
4-translation parts of a Poincar\'e transformation \cite{PT}
$(\Lambda,y)$. 
We are interested in finding the representation of the group in terms
of unitary similarity transformations, i.e.
\be
\uu a(\bbox {k},\pm)
\mapsto
e^{\pm 2i\Theta(\Lambda,\bbox k)}e^{ik\cdot y}
\uu a(\bbox {\Lambda^{-1}k},\pm)
= {\uu U}_{\Lambda,y}^{\dag} \uu a(\bbox {k},\pm){\uu U}_{\Lambda,y}\label{uno}
\ee
where $\Theta(\Lambda,\bbox k)$ is the Wigner phase. 
It is sufficient to find an appropriate representation at the
one-oscillator level. Indeed, assume we have found ${U}_{\Lambda,y}$
satisfying
\be
e^{\pm 2i\Theta(\Lambda,\bbox k)}e^{ik\cdot y} a(\bbox
{\Lambda^{-1}k},\pm) = {U}_{\Lambda,y}^{\dag} a(\bbox
{k},\pm){U}_{\Lambda,y}.
\ee
Then
$
{\uu U}_{\Lambda,y} =
\underbrace{U_{\Lambda,y}\otimes\dots\otimes U_{\Lambda,y}}_N.
$
The definition of four momentum at a single-oscillator level reads
\be
P_a &=&
\int d\Gamma(\bbox k)k_a|\bbox k\rangle\langle \bbox k|\otimes 
\frac{1}{2}\sum_s\Big(a^{\dag}_sa_s+a_s\,a^{\dag}_s\Big).
\ee
One immediately verifies that
\be
e^{iP\cdot x}a(\bbox k,s)e^{-iP\cdot x} = a(\bbox k,s) e^{-ix\cdot
k},\quad
e^{iP\cdot x}a(\bbox k,s)^{\dag}e^{-iP\cdot x} = a(\bbox
k,s)^{\dag} e^{ix\cdot k}
\ee
implying
$
U_{\bbox 1,y}=e^{iy\cdot P}.
$
Consequently, the generator of four-translations corresponding to 
$\uu U_{\bbox 1,y}=e^{iy\cdot \uu P}$ is 
$
\uu P_a=\sum_{n=1}^N P_a^{(n)}
$ 
and
\be
e^{i{\uu P}\cdot x}\uu a(\bbox k,s)^{\dag}e^{-i{\uu P}\cdot x} =
\uu a(\bbox k,s)^{\dag} e^{ix\cdot k},\quad
e^{i{\uu P}\cdot x}\uu a(\bbox k,s)e^{-i{\uu P}\cdot x} =
\uu a(\bbox k,s) e^{-ix\cdot k}.
\ee
The $x$-dependence of field operators can be introduced via 
$
{\uu F}_{ab}(x) = e^{i{\uu P}\cdot x} {\uu F}_{ab} e^{-i{\uu
P}\cdot x}
$.
Defining 
\be
U_{\Lambda,0} &=&
\exp\Big(\sum_s 2is\int d\Gamma(\bbox k)\Theta(\Lambda,\bbox k)|\bbox
k\rangle \langle \bbox k|\otimes a^{\dag}_sa_s\Big)
\Big(\int d\Gamma(\bbox p)|\bbox p\rangle
\langle \bbox {\Lambda^{-1}p}|\otimes  1\Big),
\ee
we find (\ref{uno}) with $y=0$. 
The transformations of the field tensor are finally
\be
{\uu U}_{\Lambda,0}^{\dag}{\uu F}_{ab}(x){\uu U}_{\Lambda,0} =
\Lambda{_a}{^c}\Lambda{_b}{^d}{\uu F}_{cd}(\Lambda^{-1}x),\quad
{\uu U}_{\bbox 1,y}^{\dag}{\uu F}_{ab}(x){\uu U}_{\bbox 1,y} = {\uu
F}_{ab}(x-y).
\ee
The zero-energy part of $\uu P$ can be removed by a unitary
transformation leading to a {\it vacuum picture\/} dynamics (cf.
\cite{II}). We will describe this in more detail after having
discussed the properties of states.

\section{Action of the Poincar\'e group on states}

The one-oscillator Hilbert space consists of functions $f$ satisfying
$
\sum_{n_+,n_-=0}^\infty\int d\Gamma(\bbox k)|f(\bbox k,n_+,n_-)|^2
<\infty.
$
The operator $U_{\Lambda,y}$ introduced in the previous section acts
on states of a single oscillator by
\be
|f\rangle
\mapsto
U_{\Lambda,y}|f\rangle = U_{\bbox
1,y}U_{\Lambda,0}|f\rangle
=
\sum_{n_\pm}\int d\Gamma(\bbox k)f(\bbox {\Lambda^{-1}k},n_+,n_-)
e^{2i(n_+-n_-)\Theta(\Lambda,\bbox {k})}e^{ik\cdot y(n_++n_-+1)}
|\bbox {k},n_+,n_-\rangle.\label{U1}
\ee
The Poincar\'e transformation of an arbitrary multi-oscillator state
$|\uu f\rangle$ is
$
\uu U_{\Lambda,y}|\uu f\rangle =
\underbrace{U_{\Lambda,y}\otimes\dots\otimes U_{\Lambda,y}}_N|\uu f\rangle.
$
The form (\ref{U1}) is very similar to the zero-mass spin-1
representation, the difference being in the multiplier
$n_++n_-+1$. 

In what follows we will work in a ``vacuum picture", i.e. with unitary
transformations
\be
f(\bbox k,n_+,n_-)
\mapsto
V_{\Lambda,y}f(\bbox k,n_+,n_-) &=& e^{i(n_++n_-)k\cdot y}
e^{2i(n_+-n_-)\Theta(\Lambda,\bbox {k})} f(\bbox
{\Lambda^{-1}k},n_+,n_-)\label{Uvac}.
\ee
The transition
$
U_{\Lambda,y}\mapsto V_{\Lambda,y}= W_{y}^{\dag}U_{\Lambda,y}
$
is performed by means of the unitary transformation that commutes
with reducible creation and annihilation operators.

\section{Vacuum and coherent states}

Vacuum states are all the states which are annihilated by all
annihilation operators. At the one-oscillator level these are the
states of the form
$
|O\rangle=\int d\Gamma(\bbox k)O(\bbox k)|\bbox k,0,0\rangle.
$
Even in the vacuum picture the vacuum states are not Poincar\'e
invariant since
$
V_{\Lambda,y}O(\bbox k) = O(\bbox {\Lambda^{-1}k})
$
which means they transform as a 4-translation-invariant scalar field.
We will often meet the expression $Z(\bbox k)=|O(\bbox k)|^2$
describing the probability density of the ``zero modes".

We define an $N$-oscillator vacuum state as a tensor product of $N$ copies of
single-oscillator vacua,
\be
|\uu O\rangle &=&
\underbrace{|O\rangle\otimes\dots
\otimes |O\rangle}_N.\label{uu O}
\ee
Vacuum may be regarded as a Bose-Einstein condensate of the ensemble
of harmonic oscillators at zero temperature.  Such a vacuum is
simultaneously a particular case of a coherent state with
$\alpha(\bbox k,s)=0$. 

An analogue of the standard coherent (or ``semiclassical") state is
at the 1-oscillator level
\be
|O_\alpha\rangle &=&
\sum_{n_+,n_-}\int d\Gamma(\bbox k)
O_\alpha(\bbox k,n_+,n_-) |\bbox k,n_+,n_-\rangle
\ee
where
$
O_\alpha(\bbox k,n_+,n_-) =
\frac{1}{\sqrt{n_+!n_-!}}O(\bbox k)
\alpha(\bbox k,+)^{n_+}\alpha(\bbox k,-)^{n_-}
e^{-\sum_\pm|\alpha(\bbox k,\pm)|^2/2}.
$
Under (\ref{U1}) the vacuum-picture coherent-state wave function transforms by
$
V_{\Lambda,y}O_\alpha
=
O_{T_{\Lambda,y}\alpha}
$
where
$
\alpha\mapsto T_{\Lambda,y}\alpha
$
is the usual spin-1 massless unitary representation.
Coherent states are related to the vacuum state via the displacement
operator
\be
{\uu {\cal D}}(\beta) &=& e^{\uu a(\beta)^{\dag}-\uu
a(\beta)},\quad
\uu {\cal D}(\beta)|\uu O_\alpha\rangle
= |\uu O_{\alpha+\beta}\rangle,\label{uu D}\\ 
{\uu {\cal D}}(\beta)^{\dag}\uu
a(\bbox k,s){\uu {\cal D}}(\beta) &=&
\uu a(\bbox k,s)+\beta(\bbox k,s)\uu I_{\bbox k},\quad
{\uu {\cal D}}(\beta)^{\dag}\uu I_{\bbox k}{\uu {\cal D}}(\beta) =
\uu I_{\bbox k},\label{uu D1}
\ee
as follows
$
{\uu {\cal D}}(\alpha)|\uu O\rangle =|\uu O_\alpha\rangle =
|O_{\alpha_N}\rangle\otimes\dots\otimes |O_{\alpha_N}\rangle.
$
Here $|O_{\alpha_N}\rangle$ is the 1-oscillator coherent state with
$\alpha_N(\bbox {k},s)=\alpha(\bbox {k},s)/\sqrt{N}$. The appearance
of $1/\sqrt{N}$ is of crucial importance for the
question of statistics of excitations of multi-oscillator coherent
states.
Coherent-state averages of field operators 
\be
\langle\uu O_\alpha|{^-}{\uu {\hat F}}_{ab}(x)|\uu O_\alpha\rangle
&=&
\int d\Gamma(\bbox k)e_{ab}(\bbox k)
Z(\bbox k)
\Big(\alpha(\bbox k,-)e^{-ik\cdot x}
+
\overline{\alpha(\bbox k,+)}e^{ik\cdot x}\Big)\label{1-o-cs}
\ee
are equivalent to classical electromagnetic fields. 
Let us note that (\ref{1-o-cs}) involves a ``renormalized amplitude"
$
Z(\bbox k)\alpha(\bbox k,s)=|O(\bbox k)|^2\alpha(\bbox k,s)
$
and not just $\alpha(\bbox k,s)$. The latter property is very characteristic for the reducible representation and has implications for infrared and ultraviolet regularization.

\section{Normalized multi-photon states}

Consider the vector
$
\uu a(f)^{\dag}|\uu O\rangle
$
where $\uu a(f)=\sum_s\int d\Gamma(\bbox k) \overline{f(\bbox k,s)}
\uu a(\bbox k,s)$. 
The form (\ref{uu O}) of the vacuum state implies
\be
\langle \uu O|\uu a(f)\uu a(g)^{\dag}|\uu O\rangle
=
\sum_{s}\int d\Gamma(\bbox k) Z(\bbox k)
\overline{f(\bbox k,s)}g(\bbox k,s)
=
\langle fO|gO\rangle=:\langle f|g\rangle_Z.
\ee
$fO$ denotes the pointlike product $fO(\bbox k,s)=O(\bbox k)f(\bbox
k,s)$. 
Thinking of bases in the Hilbert space one can take functions $f_i$
satisfying
$
\langle f_i|f_j\rangle_Z=\delta_{ij}\label{delta}.
$
The next theorem explains in what sense the orthogonality of
multi-photon wave packets can be characterized by the same condition
as for 1-photon states, i.e. in terms of $\langle f|g\rangle_Z$.
Denote by $\sum_\sigma$ the sum over all the permutations of the set
$\{1,\dots,m\}$.
\medskip

\noindent
{\bf Theorem 1.} Consider the vacuum state (\ref{uu O}). Then
\be
{}&{}&\lim_{N\to\infty}
\langle \uu O|\uu a(f_1)\dots \uu a(f_m)\uu a(g_1)^{\dag}\dots
\uu a(g_{m'})^{\dag}|\uu O\rangle
=
\delta_{mm'}
\sum_{\sigma}
\langle f_1|g_{\sigma(1)}\rangle_Z
\dots 
\langle f_m|g_{\sigma(m)}\rangle_Z \nonumber\\
&{}&\pp{==}=
\delta_{mm'}
\sum_{\sigma}\sum_{s_1\dots s_m}\int d\Gamma(\bbox k_1)\dots
d\Gamma(\bbox k_m)Z(\bbox k_1)\dots Z(\bbox k_m)
\overline{f_1(\bbox k_1,s_1)}\dots 
\overline{f_m(\bbox k_m,s_m)}
g_{\sigma(1)}(\bbox k_1,s_1)
\dots 
g_{\sigma(m)}(\bbox k_m,s_m)\nonumber
\ee
{\it Proof:\/} For $m\neq m'$ the scalar product is zero, which is an
immediate consequence of the fact that vacuum is annihilated by all
annihilation operators and the right-hand-side of CCR is in the
center of the algebra. So assume $m=m'$.  The scalar product of two
general unnormalized multi-photon states is
\be
{}&{}&\langle \uu O|\uu a(f_1)\dots \uu a(f_m)\uu a(g_1)^{\dag}\dots
\uu a(g_m)^{\dag}|\uu O\rangle\nonumber\\ 
&{}&=
\sum_{\sigma}\sum_{s_1\dots s_m}\int d\Gamma(\bbox k_1)\dots
d\Gamma(\bbox k_m)
\overline{f_1(\bbox k_1,s_1)}\dots
\overline{f_m(\bbox k_m,s_m)}
g_{\sigma(1)}(\bbox k_1,s_1)
\dots 
g_{\sigma(m)}(\bbox k_m,s_m)
\langle \uu O|
\uu I_{\bbox k_1}\dots \uu I_{\bbox k_m}
|\uu O\rangle\nonumber\\ &{}&=
\sum_{\sigma}\sum_{s_1\dots s_m}\int d\Gamma(\bbox k_1)\dots
d\Gamma(\bbox k_m)
\frac{1}{N^m}
\overline{f_1(\bbox k_1,s_1)}\dots
\overline{f_m(\bbox k_m,s_m)}
g_{\sigma(1)}(\bbox k_1,s_1)
\dots 
g_{\sigma(m)}(\bbox k_m,s_m)
\nonumber\\
&{}&
\times\langle \uu O|
\Big(I_{\bbox k_1}\otimes \dots\otimes I+
\dots +I\otimes\dots\otimes I_{\bbox k_1}\Big)
\dots
\Big(I_{\bbox k_m}\otimes \dots\otimes I+
\dots +I\otimes\dots\otimes I_{\bbox k_m}\Big)
|\uu O\rangle\label{multi-norm}
\ee
Further analysis of (\ref{multi-norm}) can be simplified by the
following notation:
$$
1_{k_j} = I_{\bbox k_j}\otimes \dots\otimes I;\quad
2_{k_j} = I\otimes I_{\bbox k_j}\otimes \dots\otimes I;\quad
\dots\quad N_{k_j} = I\otimes\dots\otimes I_{\bbox k_j}
$$
with $j=1,\dots,m$; the sums-integrals $\sum_{s_j}\int d\Gamma(\bbox k_j)$ are 
denoted by $\sum_{k_j}$.  Then (\ref{multi-norm}) can be
written as
\be
&{}&
\sum_\sigma
\sum_{k_1\dots k_m}
\overline{f_1(k_1)}\dots 
\overline{f_m(k_m)}
g_{\sigma(1)}(k_1)
\dots 
g_{\sigma(m)}(k_m)
\frac{1}{N^m}
\sum_{A\dots Z=1}^N
\underbrace{
\langle O|\dots\langle O|}_N
A_{k_1}\dots Z_{k_m}
\underbrace{
|O\rangle\dots |O\rangle}_N\label{az}
\ee
Since $m$ is fixed and we are interested in the limit $N\to\infty$ we
can assume that $N>m$. Each element of the sum over $A_{k_1}\dots Z_{k_m}$ 
in (\ref{az}) can be associated with a unique point
$(A,\dots,Z)$ in an $m$-dimensional lattice embedded in a cube with
edges of length $N$.

Of particular interest are those points of the cube, the coordinates
of which are all different. Let us denote the subset of such points
by $C_0$. For $(A,\dots,Z)\in C_0$
\be
\langle O|\dots\langle O|
A_{k_1}\dots Z_{k_m}
|O\rangle\dots |O\rangle = Z(\bbox k_1)\dots Z(\bbox k_m)\label{contr}
\ee
no matter what $N$ one considers and what are the numerical
components in $(A,\dots,Z)$. (This makes sense only for $N\geq m$;
otherwise $C_0$ would be empty). Therefore each element of $C_0$
produces an identical contribution (\ref{contr}) to (\ref{az}). Let
us denote the number of points in $C_0$ by $N_0$.

The sum (\ref{az}) can be now written as
\be
{}&{}&
\sum_\sigma
\sum_{k_1\dots k_m}
\overline{f_1(k_1)}\dots 
\overline{f_m(k_m)}
g_{\sigma(1)}(k_1)
\dots 
g_{\sigma(m)}(k_m) {\cal P}_0 Z(\bbox k_1)\dots Z(\bbox k_m)+\nonumber\\ 
&{}&
\sum_\sigma
\sum_{k_1\dots k_m}
\overline{f_1(k_1)}\dots 
\overline{f_m(k_m)}
g_{\sigma(1)}(k_1)
\dots 
g_{\sigma(m)}(k_m)
\frac{1}{N^m}
\sum_{(A\dots Z)\not\in C_0}
\underbrace{
\langle O|\dots\langle O|}_N
A_{k_1}\dots Z_{k_m}
\underbrace{
|O\rangle\dots |O\rangle}_N.\nonumber
\ee
The coefficient ${\cal P}_0=\frac{N_0}{N^m}$ represents a probability
of $C_0$ in the cube.  The elements of the remaining sum over
$(A\dots Z)\not\in C_0$ can be also grouped into classes according to
the values of $
\langle O|\dots\langle O|
A_{k_1}\dots Z_{k_m} |O\rangle\dots |O\rangle$. There are $m-1$ such
different classes, each class has its associated probability ${\cal
P}_j$, $0<j\leq m-1$, which will appear in the sum in an analogous
role as ${\cal P}_0$.
The proof is completed by the observation that
\be
\lim_{N\to\infty}{\cal P}_0 = 1;\quad
\lim_{N\to\infty}{\cal P}_j = 0,\quad 0<j.
\ee
Indeed, the probabilities are unchanged if one rescales the cube to
$[0,1]^m$. The probabilities are computed by means of an
$m$-dimensional uniformly distributed measure.  $N\to\infty$
corresponds to the continuum limit, and in this limit the sets of
points of which at least two coordinates are equal are of
$m$-dimensional measure zero.
\rule{5pt}{5pt}

{\it Remark\/}: The coherent-state displacement operator is given by the usual power series in multiphoton states. Since in our representation the multiphoton states behave in the thermodynamic limit as those of the usual Fock one, one expects that statistics of excitations of a coherent state is, for $N\to\infty$, Poissonian. The proof that this is indeed the case follows the lines similar to those of Theorem~1 and can be found in \cite{II}.

\section{Field operators are indeed operators}

Acting with the vector potential operator on a vacuum we obtain the
vector
\be
|{\uu A}_a(x)\rangle = {\uu A}_a(x)|\uu O\rangle=
\frac{1}{\sqrt{N}}
\Big(
|{A}_a(x)\rangle
\underbrace{
|O\rangle\dots|O\rangle}_{N-1} +\dots+
\underbrace{
|O\rangle\dots|O\rangle}_{N-1} |{A}_a(x)\rangle
\Big)
\ee
where
\be
|{A}_a(x)\rangle &=& -i\int d\Gamma(\bbox k)e^{ik\cdot x}O(\bbox
k)|\bbox k\rangle\Big( m_{a}(\bbox k) a_-^{\dag}|0,0\rangle +
\bar m_{a}(\bbox k)
a_+^{\dag}|0,0\rangle\Big).
\ee
The positive definite scalar product
$
\langle{\uu A}_a(y)|
(-g^{ab}) |{\uu A}_b(x)\rangle =
\langle{A}_a(y)|
(-g^{ab}) |{A}_b(x)\rangle = 2 \int d \Gamma(\bbox k) e^{ik\cdot
(x-y)}Z(\bbox k)
$
shows that there is no ultraviolet divergence at $x=y$ since $\int d
\Gamma(\bbox k)Z(\bbox k)=1$. 
It is easy to understand that the same property will hold also for
general states. To see this let us write the single-oscillator field
operator as a function of the operator $\hat k_a=\int d\Gamma(\bbox
k)k_a|\bbox k\rangle\langle\bbox k|$, i.e.
\be
{A}_a(x) &=& im_{a}(\hat{\bbox k})
\big(e^{-i\hat k\cdot x}\otimes a_+
-e^{i\hat k\cdot x}\otimes a_-^{\dag}\big) + i\bar m_{a}(\hat{\bbox
k})
\big(
e^{-i\hat k\cdot x}\otimes a_- -e^{i\hat k\cdot x}\otimes
a_+^{\dag}\big).
\ee
The operators $m_{a}(\hat{\bbox k})$ and $\bar m_{a}(\hat{\bbox k})$
are functions of the operator $\hat {\bbox k}$ and are defined in the
standard way via the spectral theorem. The remaining operators
($a_j$, $a_j^{\dag}$, and $e^{\pm i\hat k\cdot x}$) are also well
behaved.  Particularly striking is the fact that the {\it
distribution\/} $\int d\Gamma(\bbox k)e^{ik\cdot x}$ is replaced by
the {\it unitary\/} operator
$
e^{i\hat k\cdot x} =
\int d\Gamma(\bbox k)e^{ik\cdot x}|\bbox k\rangle\langle\bbox k|.
$
The latter property is at the very heart of various regularities
encountered in the reducible formalism.

\section{Radiation fields via $S$ matrix in reducible representation}

It is widely known \cite{IZBB,BD} that in the canonical theory the
scattering matrix corresponding to radiation fields produced by a
classical transverse current is given, up to a phase, by a
coherent-state displacement operator $e^{-i\int d^4y J(y)\cdot A_{\rm
in}(y)}$.  One of the consequences of such an approach is the
Poissonian statistics of photons emitted by classical currents. An
unwanted by-product of the construction is the infrared catastrophe.

Let us assume that we deal with a classical transverse current
$J_a(x)$ whose Fourier transform is $\tilde J_a(k)=\int d^4x
e^{ik\cdot x}J_a(x)$.  Transversality means here that
$
\tilde J_a(|\bbox k|,\bbox k)=
\bar m_a(\bbox k)\tilde J_{10'}(|\bbox k|,\bbox k) 
+ m_a(\bbox k)\tilde J_{01'}(|\bbox k|,\bbox k).
$
An interaction term representing a
classical current minimally coupled to the electromagnetic field
takes the form
$
H_{\rm int}=
\int d^3x J(x)\cdot\uu A(x)
$
and the resulting scattering matrix is
$
S=e^{i\hat\phi}e^{-i\int d^4y J(y)\cdot\uu A_{\rm in}(y)}
$
with some phase $\hat\phi$ which belongs to the center of CCR. Application of $S$ to the incoming field
produces
$
\uu A_{a\rm out}(x)
= S^{\dag}\uu A_{a\rm in}(x)S
$
and 
\be
\uu A_{a\rm rad}(x)
&=& i\int d\Gamma(\bbox k)
\uu I_{\bbox k}\Big(m_a\big(e^{-ik\cdot x}j(\bbox k,+)
-e^{ik\cdot x}\overline{j(\bbox k,-)}\big) +
\bar m_a\big(e^{-ik\cdot x}j(\bbox k,-)
-e^{ik\cdot x}
\overline{j(\bbox k,+)}\big)
\Big),\nonumber
\ee
where $m_a=m_a(\bbox k)$, and
\be
\uu a(\bbox k,s)_{\rm out}
=
\uu a(\bbox k,s)_{\rm in}+
j(\bbox k,s)\uu I_{\bbox k}\label{in-out'}=
\uu {\cal D}(j)^{\dag}
\uu a(\bbox k,s)_{\rm in}
\uu {\cal D}(j).\label{uu D2}
\ee
Comparison of (\ref{uu D2}) with (\ref{uu D1}) reveals that the $S$ matrix is in the reducible theory proportional
to the displacement operator constructed by means of the {\it reducible\/}
representation, i.e.
$
S=e^{i\hat\phi}\uu {\cal D}(j)
$
where $\hat\phi$ is in the center of CCR. 
 The average number of photons and the average four-momentum read
\be
\langle n\rangle
=
\langle \uu O_j|\uu n|\uu O_j\rangle
=
\sum_s\int d\Gamma(\bbox k)Z(\bbox k)|j(\bbox k,s)|^2,\quad
\langle P_a\rangle
=
\langle \uu O_j|\uu P_a|\uu O_j\rangle
=
\sum_s\int d\Gamma(\bbox k)k_a Z(\bbox k)|j(\bbox k,s)|^2.
\ee
A comparison with the infrared-divergent Fock result 
$
\langle n\rangle_{\rm Fock}
=
\sum_s\int d\Gamma(\bbox k)|j(\bbox k,s)|^2
$
shows that the reducible framework may regularize the divergence if 
$Z(\bbox k)\to 0$ with $\bbox k\to 0$. The point is that this is indeed
the case if the wave function vanishes at the boundary of the set of momenta. 
For massive particles this means vanishing at infinity. For massless particles the boundary contains also the origin $\bbox k=0$. This is a consequence of the fact that the cases $k=0$ and
$k\neq 0$, $k^2=0$, correspond to representations of the Poincar\'e
group induced from $SL(2,C)$ and $E(2)$, respectively.

It is quite remarkable that the ultraviolet cut-offs discussed in \cite{I} appeared automatically due to the same property of the formalism: The nontrivial structure of the vacuum state. In the
case of ultraviolet and vacuum divergences the regularization is a
consequence of square integrability of $O(\bbox k)$.

\section{Final remarks}

Let us end the paper with a few remarks on canonical commutation
relations for renormalized (for simplicity scalar) fields. It is
known that the CCR 
$
{[a(\bbox k),a(\bbox k')^{\dag}]} = \delta^{(3)}(\bbox k-\bbox
k')
$
holds in the standard formalism
only for free fields. The physical fields involve CCR in a form
$
{[a(\bbox k),a(\bbox k')^{\dag}]} = Z\delta^{(3)}(\bbox k-\bbox
k')
$
where $Z$ is related to a renormalization constant and the cut-off
$|\bbox k|<\infty$ is employed. Therefore $Z$ can be treated as a
non-zero constant only in a certain set of momenta asociated with the
cut-off. Alternatively, one can incorporate the cut-off by turning
$Z$ into a function, i.e.
\be
{[a(\bbox k),a(\bbox k')^{\dag}]} &=& Z(\bbox k)\delta^{(3)}(\bbox
k-\bbox k')\label{sclarCCR''}
\ee
and make the theory nonlocal. 
The problem with such a modification is that the action of the
Poincar\'e group
$
a(\bbox k)\mapsto U_{\Lambda,y}^{\dag}a(\bbox k)U_{\Lambda,y} =
a(\bbox{\Lambda^{-1} k})
$
will influence only the left-hand-side of (\ref{sclarCCR''}) and thus
the theory will not be Poincar\'e covariant. The reducible framework
is very close to the modification (\ref{sclarCCR''}) but with a
`quantized' $\hat Z(\bbox k)=I_{\bbox k}$ satisfying 
$
U_{\Lambda,y}^{\dag}I_{\bbox k}U_{\Lambda,y} = I_{\bbox{\Lambda^{-1}
k}}.
$
It is striking that all the results we have found for $N\to\infty$
may be summarized by the following rule: In the limit $N\to\infty$
the products of the form
$
I_{\bbox k_1}\dots I_{\bbox k_m}
$
are replaced by
$
Z(\bbox k_1)\dots Z(\bbox k_m)=|O(\bbox k_1)|^2\dots |O(\bbox
k_m)|^2.
$
In effect the reducible representation may be regarded as a covariant
implementation of the regularization (\ref{sclarCCR''}). Modification of the RHS of CCR makes our formalism nonlocal 
but in an unusual sense (to compare with other nonlocal theories cf.
\cite{Efimov,Moffat,Evens,Kleppe,Clayton,Cornish,Basu}).
An extension to a full quantum electrodynamics is a subject of ongoing study. Some preliminary results can be found in \cite{IV}.

\acknowledgments

The work was supported by Alexander von Humboldt Foundation, NATO, and University of Antwerp during my stays in Clausthal and Antwerp. 
I am indebted to I. Bia{\l}ynicki-Birula, H.-D. Doebner, H. Grosse, M. Kuna, and especially J. Naudts for critical comments at various stages of this work.


\begin{references}
\bibitem{uwaga1}Consider a charged particle rotating in a constant magnetic field. Treating the field as a parameter we obtain the frequency of the oscillation proportional to the magnitude of the field. Replacing the field by a field operator we obtain an operator frequency. The constant-field approximation implies that momenta and coordinates of the particle commute with the operator cyclotron frequency. In a more realistic case the frequency will be replaced by a frequency operator that does not commute with momenta. The example is generic in the sense that the parameters one finds in various models of harmonic oscillations typically depend on physical quantities which, at a more fundamental level, are quantized. 
The construction we give in this Letter may be regarded as a toy model, a first step towards a more realistic description where the operator frequency is not in the center of the algebra of field operators.
\bibitem{PT}In this paper when we speak of the Poincar\'e group we
mean the semidirect product of 4-translations and $SL(2,C)$, i.e. the
universal covering space of the Poincar\'e group.
\end{references}
\end{document}